\documentclass[hidelinks, 12pt]{article}

\usepackage{graphics}
 
\usepackage{natbib, bm}
\usepackage{enumerate}
\usepackage{amsmath,amssymb}
\usepackage{graphicx}
\usepackage{epsfig}
\usepackage{epstopdf}
\usepackage{rotating}
\usepackage{pdflscape}
\usepackage{multirow}
\usepackage{multicol}
\usepackage[small]{caption}
\usepackage{color}
\usepackage{mathtools}
\usepackage{subcaption}
\usepackage{longtable}

\usepackage{todonotes} 
\usepackage{comment} 

\newif\ifshow 
\showfalse 

\ifshow
  \includecomment{wrap}
\else
  \excludecomment{wrap} 
\fi

\newcommand{\nmathbf}{\bm}

\def\bfm{\nmathbf m}

\def\bfs{\nmathbf s}

\def\bfalpha  {\nmathbf \alpha}

\def\bfgamma  {\nmathbf \gamma}

\def\bfsigma  {\nmathbf \sigma}

\def\bfphi    {\nmathbf \phi}

\newcommand{\bfzero}{{\nmathbf 0}}
\newcommand{\bfone}{{\nmathbf 1}}

\def\boldfacefake#1{\kern-4pt
   \hbox{ \mathsurround=0pt
   \hbox to 0.4pt{$#1$\hss}\hbox to 0.4pt{$#1$\hss}\hbox {$#1$}}}

\newcommand{\btable}{\begin{table}[h]\centering}
\newcommand{\etable}{\end{table}}
\newcommand{\bt}{\begin{parag}\small \let\b=\nsb \let\sb=\nssb \begin{tabular}}
\newcommand{\et}{\end{tabular}\let\b=\nb \let\sb=\nsb\end{parag}}

\newenvironment{parag}{\par}{\par}

\newcommand{\be}{\begin{eqnarray}}
\newcommand{\ee}{\end{eqnarray}}
\newcommand{\ba}{\begin{eqnarray*}}
\newcommand{\ea}{\end{eqnarray*}}

\newcommand{\reals}{\mbox{\rm I\kern-.20em R}}
\newcommand{\sreals}{\mbox{\small \rm I\kern-.20em R}}

\tiny\normalsize

\newcommand{\titlefont}{\fontsize{17}{22}\selectfont\bfseries}

\begin{document}

\newpage

\title{\titlefont A Bayesian Multivariate Spatial Point Pattern Model: Application to Oral Microbiome FISH Image Data}

\author{Kyu Ha Lee$^{1, \ast}$, Brent A. Coull$^{1}$, Suman Majumder$^{2}$, \\
	Patrick J. La Riviere$^{3}$, Jessica L. Mark Welch$^{4}$, Jacqueline R. Starr$^{5,6}$\\ \\
    \textit{\small $^{1}$Harvard T. H. Chan School of Public Health, Boston, MA, U.S.A.}\\
    \textit{\small $^{2}$Indian Statistical Institute, Kolkata, WB, India}\\
    \textit{\small $^{3}$University of Chicago, Chicago, IL, U.S.A.}\\
    \textit{\small $^{4}$The Forsyth Institute, Cambridge, MA, U.S.A.}\\
    \textit{\small $^{5}$Channing Division of Network Medicine, Brigham and Women's Hospital, Boston, MA, U.S.A.}\\
    \textit{\small $^{6}$Harvard Medical School, Boston, MA, U.S.A.}\\
    {\small $^\ast$klee@hsph.harvard.edu}}

\maketitle

\begin{abstract}
\noindent  
Advances in cellular imaging technologies, especially those based on fluorescence \textit{in situ} hybridization (FISH) now allow detailed visualization of the spatial organization of human or bacterial cells. Quantifying this spatial organization is crucial for understanding the function of multicellular tissues or biofilms, with implications for human health and disease. To address the need for better methods to achieve such quantification, we propose a flexible multivariate point process model that characterizes and estimates complex spatial interactions among multiple cell types. The proposed Bayesian framework is appealing due to its unified estimation process and the ability to directly quantify uncertainty in key estimates of interest, such as those of inter-type correlation and the proportion of variance due to inter-type relationships. To ensure stable and interpretable estimation, we consider shrinkage priors for coefficients associated with latent processes that induce dependencies among point patterns. Model selection and comparison are conducted by using a deviance information criterion designed for models with latent variables, effectively balancing the risk of overfitting with that of oversimplifying key quantities. Furthermore, we develop a Bayesian hierarchical modeling approach to integrate multiple image-specific estimates from a given subject, allowing inference at both the global (across subjects) and subject-specific levels. An R package, \texttt{mspatPPM}, implements an efficient computational scheme based on Hamiltonian Monte Carlo and adaptive Metropolis-Hastings algorithms. Comprehensive numerical studies validate the reliability and practical utility of the proposed framework for model selection and for estimating quantities that characterize the multivariate spatial distribution of cell types. We apply the proposed method to microbial biofilm image data from the human tongue dorsum and find that specific taxon pairs, such as \emph{Streptococcus mitis}-\emph{Streptococcus salivarius} and \emph{Streptococcus mitis}-\emph{Veillonella}, exhibit strong positive spatial correlations, while others, such as \emph{Actinomyces}-\emph{Rothia}, show slight negative correlations. For most of the taxa, a substantial portion of spatial variance can be attributed to inter-taxon relationships.
\end{abstract}

\noindent%


\section{Introduction} \label{sec:intro}

Imaging technologies developed in the last decade are helping to create a new field of spatial biology, the study of cells and molecules in their native contexts. The  need for robust and accessible data analysis methods has been recognized among the barriers to realizing spatial biology's exciting potential \citep{atta2021computational, xiaowei2021method}. Here, we describe the development of multivariate point process models to estimate the complex dependence structures among types of cells in microscopic images. We developed Bayesian log-Gaussian Cox process (LGCP) models that offer multiple advantages over their frequentist counterparts, including straightforward quantification of uncertainty and unified parameter estimation. We applied these models to estimate relationships among different bacterial taxa in human oral biofilm images, using a meta-analytic approach to synthesize data across 100 images from five biospecimen donors.

Multi-spectral fluorescence in situ hybridization (FISH) imaging technologies enable the simultaneous identification and localization of various types of cells within a single field of view. Applied to the visualization of oral biofilms, this advanced imaging approach reveals highly structured spatial arrangements of microbial consortia ranging from tens to hundreds of microns in size. Species' spatial organization likely mediates microbial interactions that influence the overall function of the biofilm. Quantifying microbial spatial dependencies will provide clues to identify microbial interactions that affect risk of polymicrobial oral diseases \citep{dewhirst2010human, valm2011systems, mitri2011social, kim2020spatial}. 

Linear dipole analysis (LDA), as implemented in the software \texttt{daime}, is one of the most commonly used methods for analyzing spatial relationships in microbial biofilm images \citep{daims2006daime}. We sought to develop an approach that, like LDA, could be used to estimate spatial cross correlations but without its limitations: (a) LDA is capable of analyzing only two taxa at a time (bivariate); (b) the scale of the pair correlation function in LDA is restricted to the positive real values, making it less interpretable and precluding mechanisms to normalize values for comparative analyses; (c) it cannot provide uncertainty estimates for single-image analyses; it instead relies on multiple images to compute empirical standard errors, a method that becomes unreliable when the number of images is small; (d) as a non-parametric, simulation-based approach, LDA is subject to numerical instability when applied to sparse taxa; and (e) it has limited potential for extension to multi-level data structures, such as incorporating non-spatial, image-specific covariates (e.g., treatment groups or time points).

\begin{figure}[htp]
    \centering
    \begin{subfigure}[t]{0.34\textwidth}
        \centering
        \includegraphics[width=\textwidth]{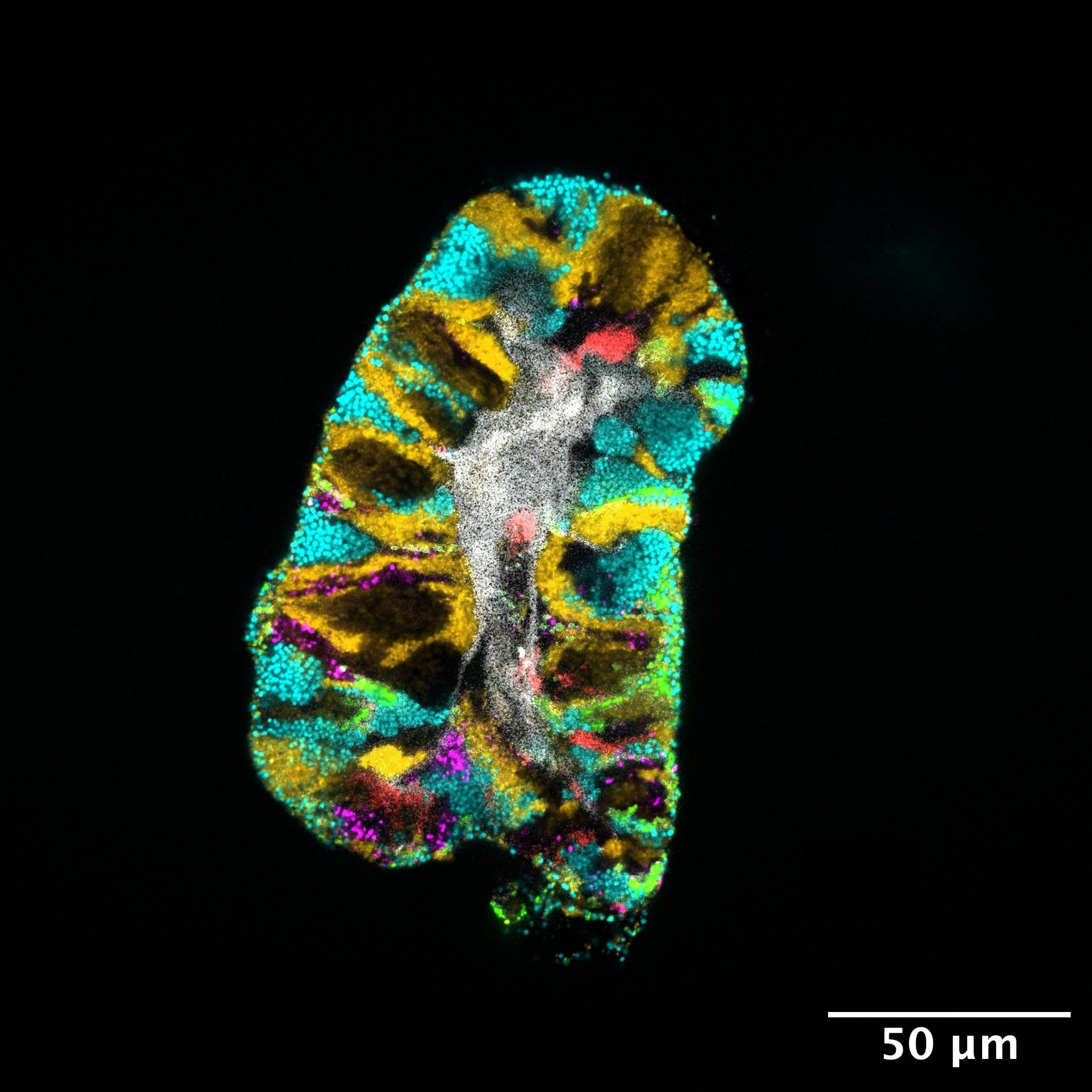}
    \end{subfigure}
    \begin{subfigure}[t]{0.34\textwidth}
        \centering
        \includegraphics[width=\textwidth]{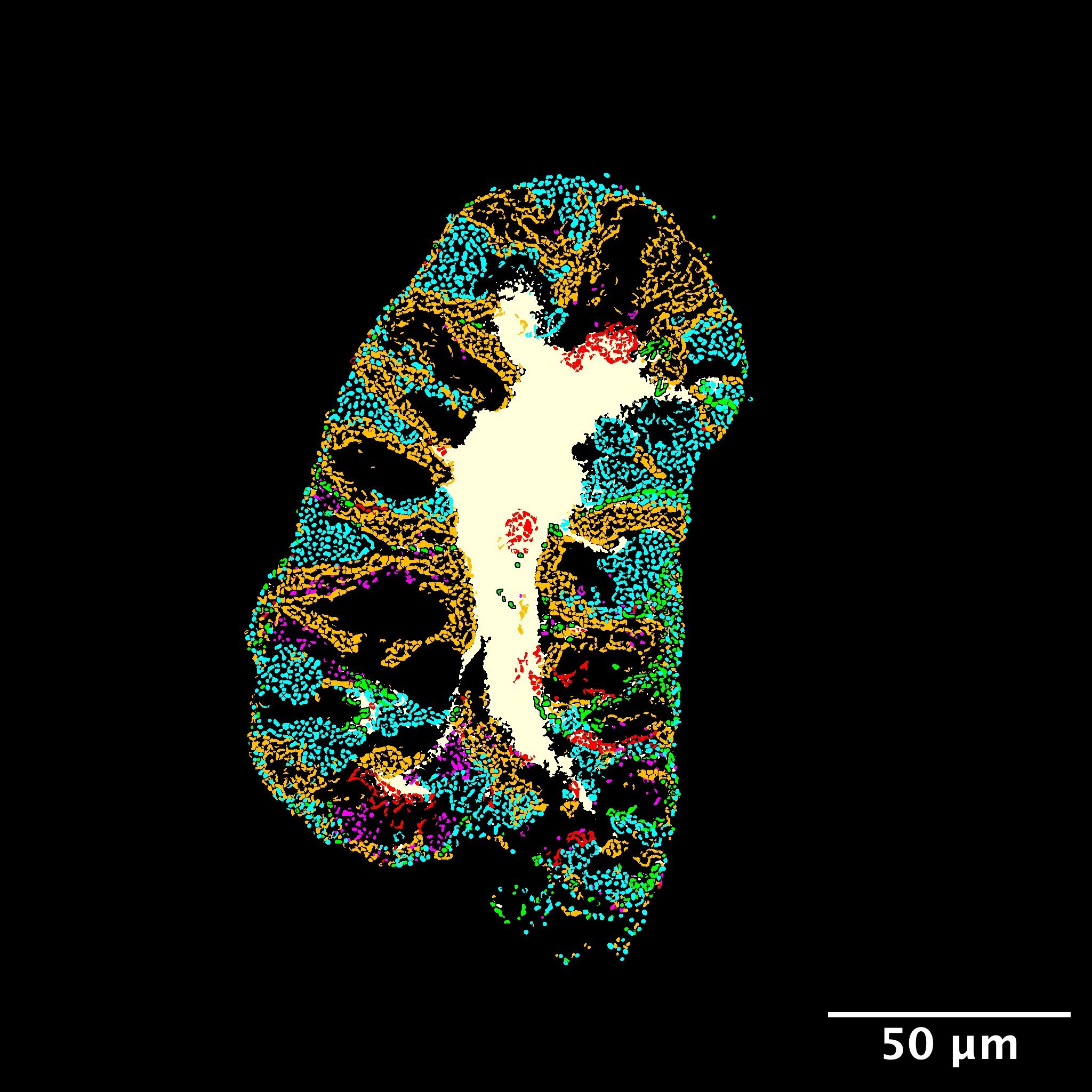}
    \end{subfigure}
    \begin{subfigure}[t]{0.29\textwidth}
        \centering
        \includegraphics[width=\textwidth]{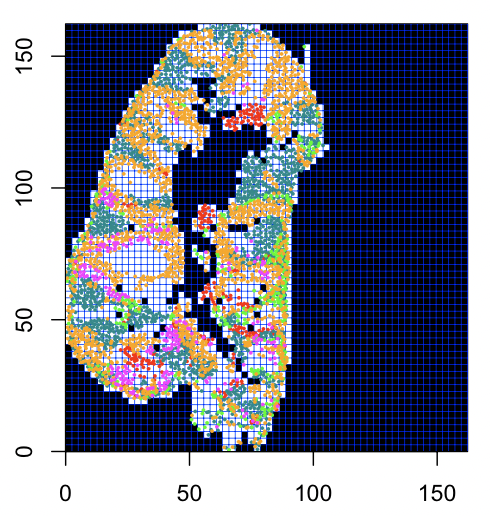}
    \end{subfigure}    
    \begin{subfigure}[t]{0.34\textwidth}
        \centering
        \includegraphics[width=\textwidth]{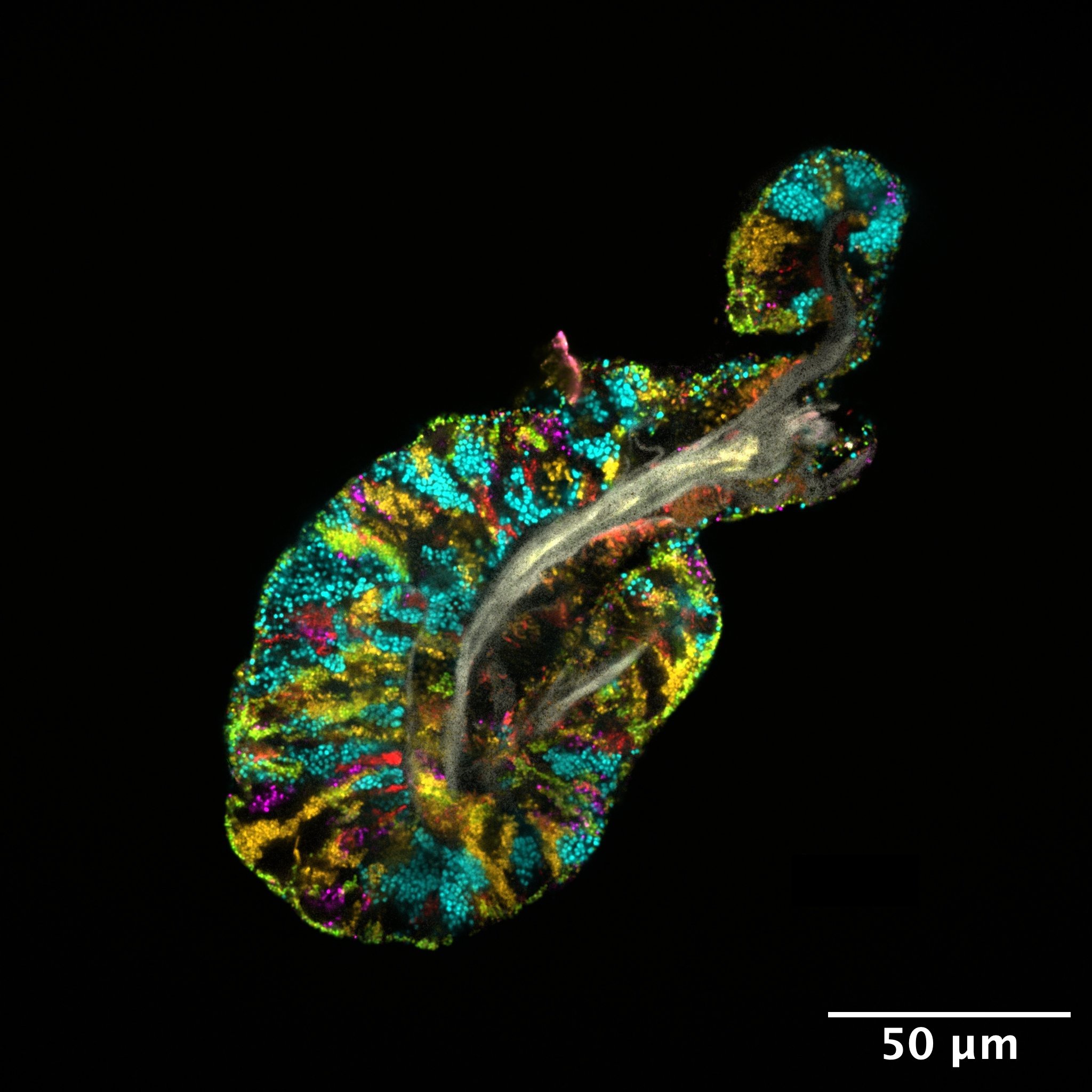}
        \caption{Unsegmented RGB image}
    \end{subfigure}
    \begin{subfigure}[t]{0.34\textwidth}
        \centering
        \includegraphics[width=\textwidth]{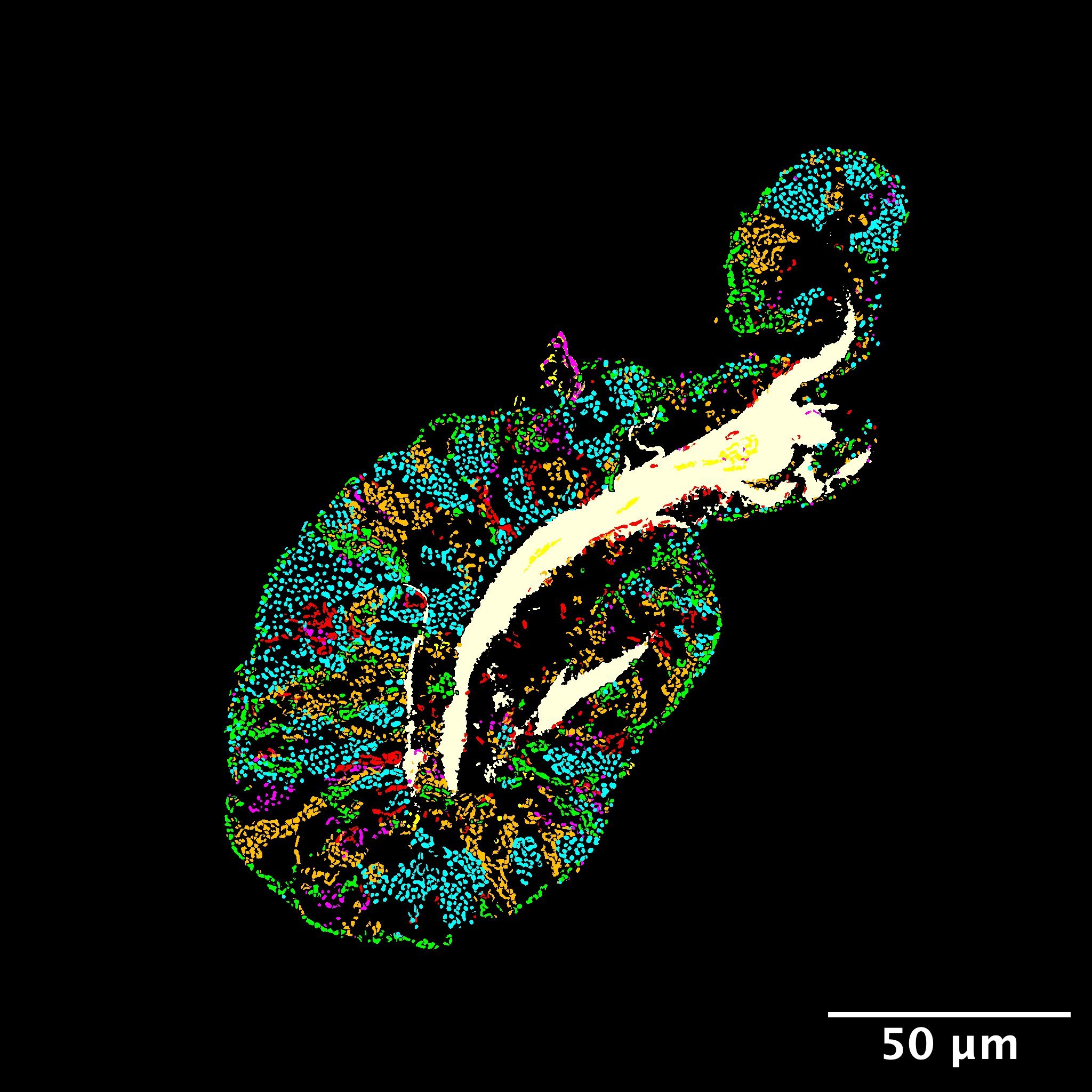}
        \caption{Segmented RGB image}
    \end{subfigure}
    \begin{subfigure}[t]{0.29\textwidth}
        \centering
        \includegraphics[width=\textwidth]{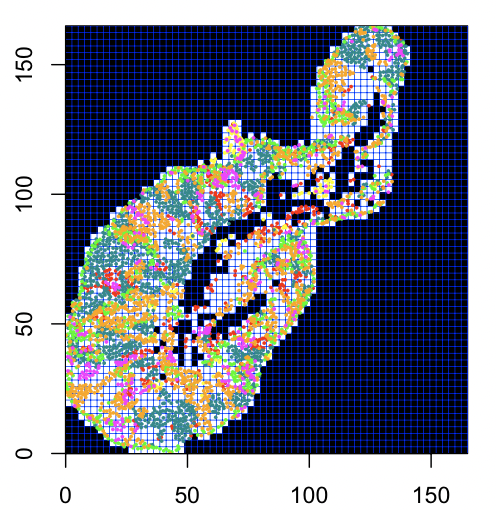}
        \caption{Points in X-Y coordinates}
    \end{subfigure}
    
    \caption{Two representative microbiome biofilm images from the human tongue dorsum, image \#3 from Subject 1 (top) and image \#11 from Subject 5 (bottom) are shown in two versions: (a) an unsegmented and (b) a segmented image. These images illustrate the spatial distribution of six bacterial taxa: \emph{Actinomyces} (Red), \emph{Rothia} (Cyan), \emph{Streptococcus mitis} (Green), \emph{Streptococcus salivarius} (Orange), \emph{Neisseriaceae} (Yellow), and \emph{Veillonella} (Magenta). Host epithelial cells, identified by autofluorescence, are displayed in white. Panel (c) presents an image generated using the coordinate information extracted from the segmented image (b) with the \texttt{R} package \texttt{mspatPPM}. For detailed summary statistics on taxa abundance and image dimensions across all image datasets, see Tables A1--A5 in Supplementary Materials A.}
    \label{fig:images_steps}
\end{figure}

A compelling way to overcome these challenges, spatial point process modeling frameworks---specifically LGCP models---are well suited to quantify spatial point patterns in which underlying intensity surfaces exhibit stochastic variation \citep{moller1998log, diggle2005point, s2019careful, samartsidis2019bayesian}. Multivariate LGCP models are designed to simultaneously handle multiple types of point processes \citep{brix2001space, diggle2013spatial, waagepetersen2016analysis, choiruddin2020regularized} such as those observed in multi-spectral FISH images. These models represent each point type (e.g., taxon) with its own intensity surface while incorporating surfaces' dependencies through shared or correlated Gaussian processes, enabling the analysis of complex interactions among point types. 

To date, most multivariate LGCP models have been developed within the frequentist paradigm, often employing a two-stage approach that separates the estimation of the intensity function from that of the covariance structure \citep{waagepetersen2016analysis}. While this framework offers a practical and efficient method for parameter estimation, it risks overlooking the uncertainty associated with the first-stage estimation during covariance structure modeling \citep{diggle1998model}. This limitation is especially problematic when the investigation of dependence structures and the prediction of point patterns are central to the analysis, as in our application. Although likelihood-based or Bayesian estimation can mitigate this issue, to the best of our knowledge, methods and software in this context have been developed only for relatively simple multivariate LGCP models with restrictive covariance structures \citep{diggle2013spatial, taylor2015bayesian} and have not yet been extended to more flexible models, such as those proposed by \cite{waagepetersen2016analysis}.

To this end, we propose a novel Bayesian framework for the spatial point pattern analysis of imaging data using a flexible multivariate LGCP model. This framework characterizes the underlying multivariate dependence structure by incorporating combinations of multiple latent processes to describe the joint distribution of multi-type processes. Compared with similarly specified frequentist models, the proposed framework provides three advantages. First, it facilitates straightforward quantification of uncertainty for estimates of interest, such as complex functions derived from model parameter estimates. This is achieved via posterior distributions, which are natural byproducts of the Markov chain Monte Carlo (MCMC) sampling process. In contrast, frequentist methods often rely on parametric bootstrap simulations from the fitted model, which can underestimate the variance of parameter estimates \citep{waagepetersen2016analysis}. Second, parameter estimation is performed through a unified MCMC sampling scheme, avoiding the risk of underestimating covariance structure parameters---another common limitation of the conventional, frequentist two-stage approach. Third, the number of distinct lags does not need to be pre-specified for model fitting, as is required in existing two-stage estimation methods.

The proposed framework incorporates several innovative components designed for the Bayesian paradigm, addressing challenges related to model fitting, model selection, and the synthesis of estimates across datasets. To support robust model fitting, we improve numerical stabilization in high-dimensional models by employing lasso and weakly informative normal priors for the coefficients associated with the latent processes that induce dependence among multi-type point patterns. For effective model selection, we use a deviance information criterion (DIC) specifically devised for models with latent variables, balancing the risks of overfitting and oversimplifying the dependence structure. Finally, to enable the synthesis of estimates across multi-level imaging data, we develop a Bayesian hierarchical modeling approach that integrates image-specific estimates and facilitates inference at both the global and subject-specific levels. All computational algorithms, including the model fitting and post-fitting framework, are implemented in the R package \texttt{mspatPPM}. In this paper, we present a comprehensive description of the proposed framework, assess its performance through simulation studies, and highlight its practical utility with application to real datasets.

\section{Microbial Biofilm Image Data from the Human Tongue Dorsum}\label{sec:data}

The model development and data analysis in this work were motivated by previously published image data displaying the spatial organization of the tongue microbiome. In that study, microbiota from the human tongue dorsum were collected by using a ridged plastic tongue scraper, a method that preserves the biofilm's spatial structure. Samples were fixed in ethanol and spread onto slides for multiplexed FISH imaging. The samples contained discrete bacterial biofilms, termed ``consortia," that were characterized by dense layers of bacteria surrounding a core of human epithelial cells. From one slide for each subject, 20 large consortia were selected for further analysis. Detailed data collection methods have been described previously \citep{wilbert2020spatial}. The FISH probes employed in this study targeted 1 phylum, 17 genera, and 7 species within these genera. Taxa present in all subjects and in $\geq$95\% of images acquired were \emph{Actinomyces}, \emph{Rothia}, \emph{Veillonella}, \emph{Neisseriaceae}, and \emph{Streptococcus}. Our analysis focuses on three genera (\emph{Actinomyces}, \emph{Rothia}, and \emph{Veillonella}) and two species (\emph{Streptococcus mitis} and \emph{Streptococcus salivarius}), excluding \emph{Neisseriaceae} due to its relative sparseness.

In the previous analysis, the biofilm images were segmented using \texttt{Fiji} \citep{schindelin2012fiji} by applying a 3 $\times$ 3 median filter followed by the ``Auto Local" thresholding function using the Bernsen method (Figure \ref{fig:images_steps}b; details available on the ImageJ website: https://imagej.net/plugins/auto-local-threshold). Spatial coordinates for each cell's centroid were generated using the ``Analyze Particles" function in \texttt{Fiji}, with a size filter set to a minimum diameter of 0.5 $\mu m$. After removing regions of the image without observations, the bottom-left corner of the image was assigned the coordinate $(0,0)$ (Figure \ref{fig:images_steps}c).
 
From the original set of 100 images (20 images per subject across 5 subjects), we created 14 additional images by subdividing those containing multiple consortia. Subsequently, a different set of 14 images were excluded due to insufficient counts ($<$10) for any given taxon, leaving a total of 100 images included in further analyses (Tables A1 through A5 of Supplementary Materials A).

\section{Bayesian Framework for Multivariate Spatial Point Pattern Analysis}\label{sec:method}

\subsection{Multivariate log-Gaussian Cox process model} \label{sec:mLGCP}

Let $J$ denote the number of point types (e.g., bacterial taxa in this application or---for FISH images more broadly---cell types) observed, and let $K$ represent the number of shared latent processes used to model spatial dependence across types. We consider a multivariate process $\mathbf{Z}=(Z_1, \ldots, Z_J)$, where each $Z_j$ ($j=1,\ldots,J$) is an inhomogeneous Poisson process characterized by a random intensity function $\lambda_j(\mathbf{s})$, with $\mathbf{s}\in S\subseteq\mathbb{R}^2$, where $S$ is a bounded planar window. The process $Z_j$ is assumed to be independent, conditional on the intensity function $\lambda_j(\mathbf{s})$. We further consider the following specification of the intensity functions \citep{waagepetersen2016analysis}:
\be
\lambda_j(\mathbf{s}) = \exp\left(m_j + U_j(\mathbf{s}) + \sum_{k=1}^K \alpha_{jk} U_{0k}(\mathbf{s})\right), \nonumber
\ee
where $m_j$ represents the overall mean of the intensity process; $U_j(\mathbf{s})$ and $U_{0k}(\mathbf{s})$, $j=1,\ldots,J$, $k=1,\ldots,K$, are zero-mean Gaussian processes; and $\alpha_{jk}$ is the coefficient that determines the contribution of the latent process $U_{0k}(\mathbf{s})$ to $\lambda_j(\mathbf{s})$. We complete the specification of the processes $U_j(\mathbf{s})$ and $U_{0k}(\mathbf{s})$ using exponential covariance functions, which are special cases of two widely used covariance function families: the power exponential and the Mat\'{e}rn covariance functions. Specifically, the covariance functions are defined as $\sigma_j^2\exp(-\phi_j^{-1}||\bfs-\bfs'||)$ and $\sigma_{0k}^2\exp(-\phi_{0k}^{-1}||\bfs-\bfs'||)$, where $\sigma_j^2$ and $\sigma_{0k}^2$ represent the marginal variances for $U_j(\mathbf{s})$ and $U_{0k}(\mathbf{s})$, respectively, and $\phi_j$ and $\phi_{0k}$ are the spatial range parameters that control the rate of spatial correlation decay for the same processes, respectively. To ensure the identifiability, we set $\sigma_{0k}^2=1$.

While the Gaussian processes $U_j(\mathbf{s})$ and $U_{0k}(\mathbf{s})$ are assumed to be mutually independent, the resulting intensity functions $\lambda_j(\mathbf{s})$'s are not independent due to the latent structure shared through the $U_{0k}(\mathbf{s})$'s. Therefore, $ \mathbf{Z} $ can be viewed as a multivariate extension of the LGCP model, arising from correlated latent fields \citep{diggle2013spatial, waagepetersen2016analysis}.

\subsection{The observed data likelihood}

Suppose we observe a spatial pattern from $G_j$ points for type $j$, denoted by $\vec{\bfs}_j = \{\bfs_{j1},\ldots,\bfs_{jG_j}\}$, within $S$ with area $|S|$. Using the density of a Poisson process expressed based on the Radon-Nikodym derivative \citep{moller1998log}, the likelihood contribution of observed point pattern data $\vec{\bfs}_j$ can be written as 
\be
	\pi(\vec{\bfs}_j | \lambda_j(\bfs))&=& \exp\left\{|S|-\int_S \lambda_j(\bfs) d\bfs\right\}\prod_{g=1}^{G_j}\lambda_j(\bfs_{jg}). \label{eq:likelihood}
\ee
Parameter estimation within likelihood-based frequentist or Bayesian inference methods requires the explicit specification of the likelihood function (\ref{eq:likelihood}). A common approach is to construct a computational grid over the spatial domain, consisting of an $M \times N$ uniform grid of equally spaced cells, and to characterize $\log\{\lambda_j(\mathbf{s})\}$ by its value at the centroid of each cell, $\mathbf{c}_i\in S\subseteq\mathbb{R}^2$, for $i=1,\ldots,MN$, under the assumption of constant $\log\{\lambda_j(\mathbf{s})\}$ within each cell. Given this computational grid, the unique log-intensity values, denoted by $\mathbf{Y}_j=(Y_{j,1}, \ldots, Y_{j,MN})^{\top}$, follow a multivariate normal (MVN) distribution:
\be
\mathbf{Y}_j \sim \text{MVN}\left(m_j \mathbf{1}_{MN}, \mathbf{\Sigma}_j(\phi_j, \phi_0, \sigma_j^2, \bfalpha_j)\right), \label{eqn:lgcpMVN_ori}
\ee
where $\mathbf{1}_{MN}$ is an $MN \times 1$ vector of ones, $\bfalpha_j=(\alpha_{j1}, \ldots, \alpha_{jK})^{\top}$, and $\mathbf{\Sigma}_j(\Theta)$ is the $MN \times MN$ covariance matrix parameterized by $\Theta$. With the choice of covariance functions outlined in Section \ref{sec:mLGCP}, the $(i, i')$-th entry of $\mathbf{\Sigma}_j(\phi_j, \phi_0, \sigma_j^2, \bfalpha_j)$ is given by $\sigma_j^2 \exp\left(-\phi_j^{-1} \|\mathbf{c}_i - \mathbf{c}_{i'}\| \right) + \sum_{k=1}^K \alpha_{jk}^2 \exp\left(-\phi_{0k}^{-1} \|\mathbf{c}_i - \mathbf{c}_{i'}\| \right)$, for $i, i'=1,\ldots,MN$.

\subsection{Quantification of the multivariate dependence structure}
As mentioned above, one objective regarding microbial biofilm image data is to quantify and estimate between-type (inter-taxon) relationships. Another objective is to estimate the proportion of variance explained by between-type (compared with within-type) relationships. We achieve these goals by using quantities derived from moments of the multivariate LGCP models \citep{waagepetersen2016analysis, choiruddin2020regularized}. The intensity function of $Z_j$ is given by $\rho_j(\bfs)=\exp\left(\mu_j + \sigma_j^2/2 + \sum_{k=1}^K\alpha_{jk}^2/2 \right)$ and the cross pair correlation function for $Z_j$ and $Z_{j'}$ at distance $d=||\bfs-\bfs'||$ by
\be
    g_{jj'}(d)= \exp\left\{I(j=j')\sigma_j^2\exp(-\phi_j^{-1}d) + \sum_{k=1}^K\alpha_{jk}\alpha_{j'k}\exp(-\phi_{0k}^{-1}d)\right\}. \label{eq:cpf}
\ee
The second term of (\ref{eq:cpf}) represents the covariance due to the latent fields shared across different types at distance $d$, thereby inducing inter-type dependence. We refer to this term as the ``inter-type" covariance, defined as $\psi_{jj'}(d)=\sum_{k=1}^K\alpha_{jk}\alpha_{j'k}\exp(-\phi_{0k}^{-1}d)$.

The conditional intensity function of $Z_j$, given the presence of a point from $Z_{j'}$ at location $\bfs'$, is given by $\rho_j(\bfs)g_{jj'}(||\bfs-\bfs'||)$ \citep{coeurjolly2017tutorial, choiruddin2020regularized}. An ``attractive" relationship between types $j$ and $j'$, where the presence of a point in $Z_j$ at $\bfs$ is associated with increased intensity of $Z_{j'}$ at $\bfs'$, is indicated by $g_{jj'}(||\bfs - \bfs'||) > 1$ or equivalently $\psi_{jj'}(||\bfs - \bfs'||)>1$. Conversely, $g_{jj'}(||\bfs - \bfs'||) < 1$, and therefore $\psi_{jj'}(||\bfs - \bfs'||)<1$, may support a ``repulsive" relationship between two types. 

Since $g_{jj'}(d)$ and $\psi_{j,j'}(d)$ are real-valued functions, their absolute scales can make interpretation challenging. To provide a standardized measure of inter-type association, we define the inter-type cross correlation function (CCF):
\be
CCF_{jj'}(d)=\psi_{jj'}(d)/\sqrt{\psi_{jj}(0)\psi_{j'j'}(0)}, \label{eqn:ccf_reduced}
\ee
which is normalized to the interval $[-1,1]$. This rescaling facilitates interpretation. Parametric LGCP models also provide a means to quantify the proportion of inter-type variance relative to the total variance (PV) at a given distance $d$, which is given by:
\be 
        PV_j(d) = \frac{\psi_{jj}(d)}{\log \{g_{jj}(d)\}}. \label{eqn:pv}
\ee

\subsection{Prior distributions}

To fully specify the proposed Bayesian framework, we assign prior distributions to the model parameters. Frequentist approaches for high-dimensional multivariate LGCP models (i.e. large $J$) have employed regularization techniques such as the elastic net penalty for the coefficients of the latent Gaussian processes, i.e. the factor loading parameters $\alpha_{j k}$ to achieve sparsity \citep{choiruddin2020regularized}. However, even in lower dimensional models as in the oral biofilm image data ($J=5$), the computational grid representation gives rise to a high-dimensional parameter space, for example, $MN\times MN$ covariance matrices for each latent process.  

To achieve stable and interpretable model estimation, we adopt Bayesian lasso priors \citep{park2008bayesian} for $\bfalpha_j$, which are key parameters involved in both $C_{jj'}(d)$ and $PV_j(d)$. Shrinkage priors, like the Bayesian lasso, have been shown to improve model fit and predictive power across various models and data scenarios, in both low- and high-dimensional settings \citep{park2008bayesian, lee2011bayesian, lee2015survival, lee2017variable, reeder2024group}. Within the Bayesian paradigm, the lasso estimate corresponds to the posterior mode under independent Laplace priors for the model parameters. 

Since the Laplace distribution can be expressed as a scale-mixture of normals with a gamma mixing density \citep{andrews1974scale}, we use the following computationally convenient representation of Bayesian lasso priors for $\alpha_{j k}$:
\be
	\alpha_{j k}|\tau_{j k}^2 &\sim& \textrm{Normal}(0, ~h_{\alpha}\tau_{j k}^2), \nonumber \\
    \tau_{j k}^2 | \lambda^2 &\sim& \textrm{Gamma}(1, ~\lambda^2), \label{eqn:BL}
\ee
where $\lambda$ is a tuning parameter that controls the degree of shrinkage applied to the posterior distribution of the $\alpha_{j k}$. Unlike conventional normal priors, $\alpha_{j k}\sim\textrm{Normal}(0, ~h_{\alpha})$, the prior variance under (\ref{eqn:BL}) is adaptively adjusted and estimated through the tuning parameter. 

For the mean of the log-intensity processes and the variance of $U_j(\bfs)$, we use $m_j \sim \textrm{Normal}(\mu_m, ~h_{m})$ and $\sigma_{j} \sim \textrm{Gamma}(a_{\sigma}, ~b_{\sigma})$. Uniform priors are assigned on the log-transformed spatial range parameters: $\log\phi_j$ $\sim$ Uniform($l_{\phi}$, $u_{\phi}$) and $\log\phi_{0k}$ $\sim$ Uniform($l_{\phi0}$, $u_{\phi0}$). The hyperparameters ($a_{\lambda}$, $b_{\lambda}$, $h_{\alpha}$, $\mu_{m}$, $h_m$, $a_{\sigma}$, $b_{\sigma}$, $l_{\phi}$, $u_{\phi}$, $l_{\phi0}$, $u_{\phi0}$) need to be prespecified.

\section{Computational Scheme}\label{sec:comp}

Estimation and inference for the proposed Bayesian framework are performed based on samples from the joint posterior distribution either by exploiting prior-posterior conjugacies or by using an adaptive MCMC algorithm. However, MCMC sampling is challenged by the inversion and Cholesky factorization of the high-dimensional matrix $\mathbf{\Sigma}_j$ in (\ref{eqn:lgcpMVN_ori}), which is of size $MN\times MN$.

To overcome the resulting computational burden, we use numerical techniques based on the 2D fast Fourier transform (FFT) \citep{moller1998log}. First, we extend the original $M\times N$ grid to an $M_e\times N_e$ grid and wrap it on a torus, where $M_e=2^g$, $N_e=2^g$, and $g$ is an integer such that $M_e\geq 2(M-1)$, $N_e\geq 2(N-1)$. Then a block-Toeplitz correlation matrix derived from $\mathbf{\Sigma}_j$ is extended to a block-circulant matrix \citep{wood1994simulation}. The computational challenges associated with high-dimensional covariance matrices, including operations such as inversion and square root can be significantly alleviated by reformulating them as matrix-vector and element-wise matrix multiplications using the FFT \citep{rue2005gaussian, teng2017bayesian}. Further, we adopt the following reparametrization of $\mathbf{Y}_j$:
\be
	\mathbf{Y}_j = m_j \mathbf{1}_{MN} + \sigma_j C^{1/2}_j \bfgamma_j + \sum_{k=1}^K \alpha_{jk}C^{1/2}_{0k}\bfgamma_{0k}, \label{eqn:lgcpMVN_repara}
\ee
where $\bfgamma_j$ and $\bfgamma_{0k}$ follow MVN($\bfzero$, $I_{MN}$). The $(i, i')$-th entry of correlation matrices $C_j$ and $C_{0k}$ are defined as $\exp\left(-\phi_j^{-1} \|\mathbf{c}_i - \mathbf{c}_{i'}\| \right)$ and $\exp\left(-\phi_{0k}^{-1} \|\mathbf{c}_i - \mathbf{c}_{i'}\| \right)$, respectively, for $i, i'=1,\ldots,MN$. This parametrization separates the correlation matrix from the variance and latent variables $\bfgamma_j$ and $\bfgamma_{0k}$, allowing complex matrix operations to be avoided in the updates of the other components and thereby improving computational efficiency. 

There are two types of approaches to estimate the tuning parameter $\lambda$ in the Bayesian lasso: empirical Bayes and full Bayes. The Monte Carlo expectation-maximization steps in the empirical Bayes approach will introduce additional computational burden, as they require calculating the expected mean and variance of the tuning parameters, which in turn involves simulating high-dimensional models within each MCMC iteration. Therefore, we adopt the full Bayes approach, assigning the gamma hyperprior, $\lambda^2 \sim \textrm{Gamma}(a_{\lambda}, ~b_{\lambda})$, and update the parameter in a manner analogous to other model parameters.

We propose a novel MCMC sampler combining the Hamiltonian Monte Carlo algorithm \citep{neal2012bayesian} to efficiently sample the high-dimensional parameters, $\bfgamma_j$ and $\bfgamma_{0k}$, with a Hessian-based Metropolis-Hastings algorithm \citep{chib1994bayes, geweke2003note} for the other parameters (details in Supplementary Materials B). After evaluating several software options for performing FFT, including our custom implementation that uses the external \texttt{FFTW} library \citep{frigo2005design}, we find that the built-in R functions, written in \texttt{C}, offer the highest efficiency and thus form the basis of our algorithm. To facilitate the efficient management of multiple image datasets, we develop custom \texttt{C} functions for constructing the computational grid. These are integrated into the software package, further enhancing computational performance. The complete algorithm is available on the first author's GitHub page (https://github.com/kleeST0/mspatPPM). Running 10,000 MCMC iterations for the microbial biofilm image data described here takes approximately 9 minutes on an Apple M2 Max MacBook Pro. This timing corresponds to the model with $K$=4 on the extended grid, where $M_e=N_e=128$.

\section{Post-Model Fitting Framework}

\subsection{Model selection and evaluation} \label{sec:dic}
The proposed Bayesian multivariate LGCP model requires identifying an appropriate number of latent processes ($K$). An unnecessarily large $K$ increases the risk of overfitting, while a too-small $K$ may result in oversimplified estimates of key quantities, such as the spatial cross correlation coefficients or PV. \cite{waagepetersen2016analysis} discussed identifiability conditions for the latent process model and noted that, although large values of $K$ are not ruled out theoretically, practical concerns lead to a recommendation of restricting $K\leq J$. Since we adopt the same latent model structure, we follow the same guideline. To perform this critical task, we use the DIC as a model comparison metric, which balances model complexity with its ability to fit the data. Specifically, we consider a version of the DIC designed for mixtures of distributions and random effect models, DIC$_3$ in \cite{celeux2006deviation}. 

Let $y_{ij}$ and $n_{ij}$ represent the log-intensity and the observed count of points for type $j$ in the $i$-th extended computational grid cell, respectively, where the area of the grid cell is $A_i$, as described in Section \ref{sec:comp}. Let $f_{pois}(x|a)$ denote the Poisson density function with rate $a$ evaluated at $x$. Then, under the proposed model, the DIC can be estimated by using the following Monte Carlo approximation: 
\be
    \widehat{\mbox{DIC}}=-\frac{4}{R}\sum_{r=1}^R\log\left\{\prod_{j=1}^J\prod_{i=1}^{M_eN_e} f_{pois}\left(n_{ij}| A_i y_{ij}^{(r)}\right)\right\}\ +\ 2\log\left\{ \prod_{j=1}^J\prod_{i=1}^{M_eN_e}\frac{1}{R}\sum_{r=1}^R f_{pois}\left(n_{ij}| A_i y_{ij}^{(r)}\right)\right\},\nonumber
\ee
where $y_{ij}^{(r)}$ represents the value of $y_{ij}$ at the $r$-th MCMC iteration, $r=1,\ldots, R$. 

In our numerical studies, we use the $\widehat{\mbox{DIC}}$ to select $K$ and to compare models with two different prior distributions for $\alpha_{jk}$, the Bayesian lasso and normal priors. A general rule of thumb for model comparison suggests that differences of less than 2 are negligible, differences between 2 and 6 indicate positive support for the model with the lower value, and differences greater than 6 provide strong support for the model with the lower value \citep{spiegelhalter2002bayesian}.

\subsection{Synthesizing findings from multi-level image data} \label{sec:pooled}

The proposed multivariate LGCP model enables us to quantify spatial relationships in  each image. There remains a need for a robust framework to synthesize image-specific quantities across multiple images from multiple subjects, i.e., multi-level image data.

Let $z_{ml}$ denote either the Fisher z-transformed inter-type correlation for a given pair or the logit-transformed PV for a type (taxon) at a specified distance, estimated from the analysis of the image $m=1,\ldots, n_l$ of subject $l=1,\ldots, L$. To estimate both the global and subject-specific parameters, a pooled analysis approach based on Bayesian multilevel modeling is developed (described in detail in Supplementary Materials C, which includes supporting simulation studies). The algorithm is publicly available through the \texttt{R} package \texttt{mspatPPM}.

The Bayesian multi-level modeling framework is appealing as it directly provides full posterior distributions for subject-specific estimates through the conditional model, allowing for straightforward quantification of associated uncertainties \citep{dominici2000combining, coull2003bayesian}. This conditional model contrasts with the marginal models commonly used in the frequentist paradigm, which focus on the variance of the random effects rather than on the random effects themselves. Furthermore, the shrinkage effects introduced through priors can help stabilize parameter estimation, especially with a small number of subjects ($L=5$) as in our application.

\section{Simulation Studies}

We conducted simulation studies to compare the performance of the proposed framework using the Bayesian lasso prior as defined in (\ref{eqn:BL}) and the conventional $\textrm{Normal}(0, ~h_{\alpha})$ prior for $\alpha_{jk}$. We estimated CCFs and PVs based on the simulation setup described in \cite{waagepetersen2016analysis}. These models are referred to as LGCP-BL and LGCP-N, respectively. We assessed the effectiveness of the DIC metric for model selection.

\subsection{Set-up and data generation} \label{sec:simSetup}

We considered two different scenarios varying the true number of shared latent processes ($K$). Data on the unit square were simulated under the model described in Section \ref{sec:mLGCP}, with $J$=$5$ types. 

\begin{figure}[htp]
\centering
\includegraphics[width=15.5cm]{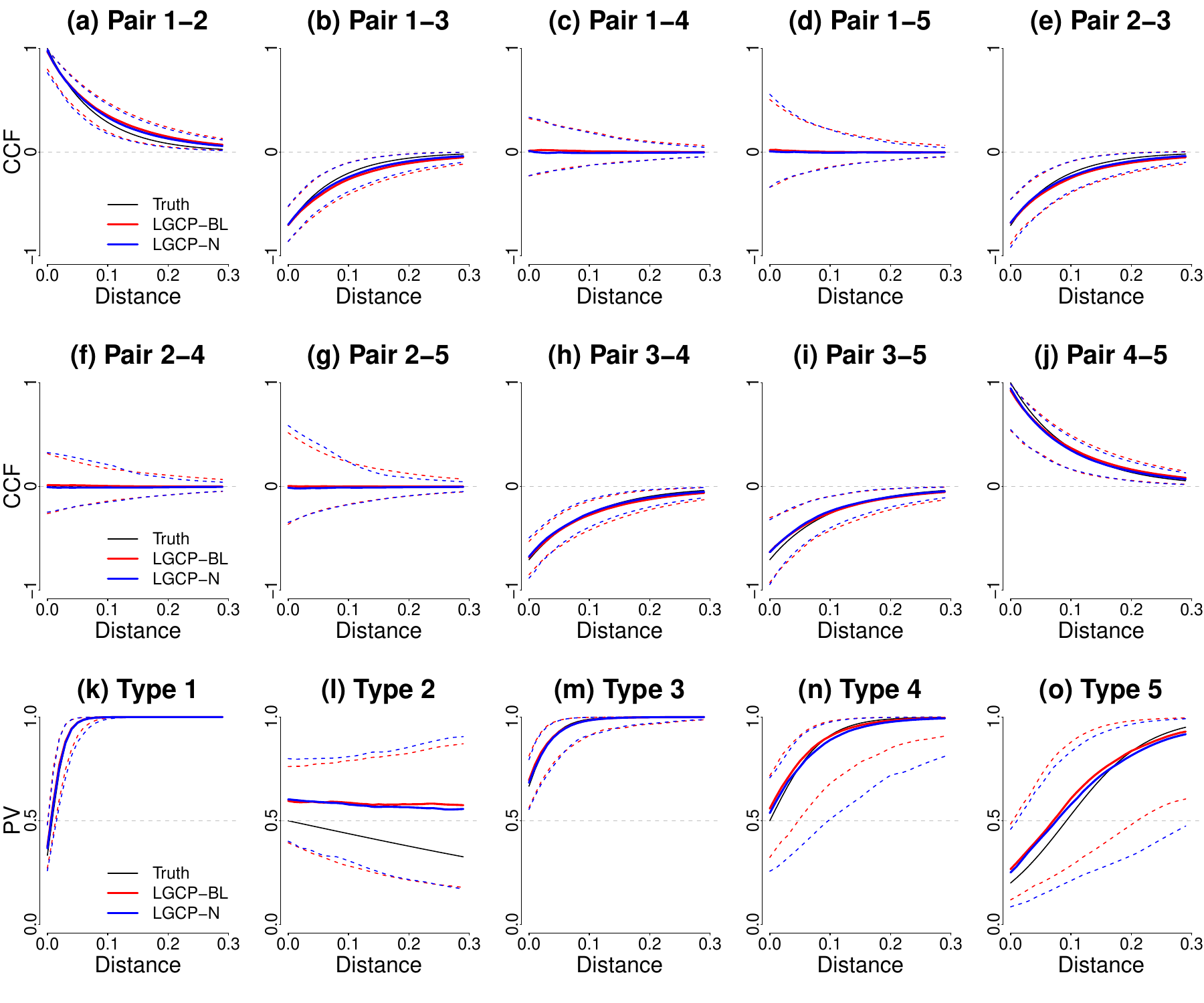}
\caption{Scenario I: the median (solid curves), 2.5\%, and 97.5\% (dashed curves) quantiles of the pointwise posterior medians for cross correlation functions (CCFs) for 10 pairs of types (a)-(l) and proportion of variances (PVs) (k)-(o). The results were summarized based on the best model selected by comparing the deviance information criterion (DIC) across the 300 simulated datasets.} \label{fig:sim_ccf_pv_sel}
\end{figure}

In Scenario I, we assumed that $K$=$2$ shared latent processes characterized the dependence structure among the five point patterns. The parameter settings were as follows: $\bfsigma=[\sigma_j]_{j=1}^J = \bfone_5$, $\bfphi=[\phi_j]_{j=1}^J = (0.01, 0.1, 0.02, 0.03, 0.04)^{\top}$,  $\bfphi_0=[\phi_{0k}]_{k=1}^K = (0.08, 0.1)^{\top}$, and $\textrm{vec}(\bfalpha)=(\sqrt{0.5}, 0, 1, 0, -1, 1, 0, -1, 0, -0.5)^{\top}$, where $\bfalpha$ is a $K\times J$ matrix with $(k,j)$-th element $\alpha_{jk}$. $\bfm=[m_j]_{j=1}^J$ was adjusted such that the expected number of points was 1,000. This setup is the same as that of \cite{waagepetersen2016analysis}, except that we set ($\phi_{2}, \phi_{01})=(0.1, 0.08)$ instead of ($0.02$, $0.02$). This modification was primarily designed to introduce variation in PV$(d)$, resulting in a mix of increasing and decreasing trends (Figure \ref{fig:sim_ccf_pv_sel}(k)-(o)). This contrasts with their original settings, which produced strictly non-decreasing trends over $d$ for all five types. Note that the resulting PV(0) and CCF(0) evaluated in \cite{waagepetersen2016analysis} remain identical to ours, as $\bfphi$ and $\bfphi_0$ do not influence these quantities at $d=0$. This alignment allowed us to use their results as a reasonable benchmark for comparison.

In Scenario II, the dependence structure among the five point patterns were constructed using $K$=$3$ shared latent processes. The parameter settings were the same as those in the first scenario except $\bfphi_0=(0.08, 0.1, 0.05)^{\top}$, and $\textrm{vec}(\bfalpha)$=($\sqrt{0.5}$, 0, 0, $-1$, 1, 0, 0, $\sqrt{0.5}$, 0, 0, 0, $-\sqrt{0.5}$, 1, $-1$, 1)$^{\top}$. This setting allows evaluation of the model's capture of a diverse combination of trends in CCF and PV functions, similar to Scenario I. 

\subsection{Specification of hyperparameters and analysis settings}\label{sec:sim_hyperSetting}

The proposed LGCP-BL and LGCP-N were fitted to 600 simulated datasets, 300 under each of the two scenarios described in Section \ref{sec:simSetup}. The proposed framework requires the specification of hyperparameters, the selection of computational grid resolution, and the choice of the number of shared latent processes ($K$).

The hyperparameters $(\mu_m, h_m)$ were set to $(0, 10^5)$ for a non-informative prior for the $\bfm$. For the remaining parameters, weakly informative priors were used to regularize those known to have little information provided by the data for estimation and to enhance numerical stability \citep{moller2003statistical, diggle2013spatial, taylor2015bayesian}. Specifically, we set $(a_\sigma, b_\sigma) = (10, 10)$, which corresponds to an induced prior distribution for $\sigma^2$ with a median of 0.94 and with 95\% of the central mass between 0.23 and 2.92. As noted in \cite{moller2003statistical}, non-informative priors are not feasible for $\log(\phi_j)$ and $\log(\phi_{0k})$, and eliciting highly informative priors for these parameters is similarly challenging. Additionally, unbounded priors for spatial range parameters are discouraged due to numerical stability issues. For example, circulant embedding techniques, described in Section \ref{sec:comp}, often fail when parameter values exceed thresholds determined by the extended grid size \citep{graham2018analysis}. To address these concerns, we chose hyperparameters $l_{\phi}$, $u_{\phi}$, $l_{\phi0}$, and $u_{\phi0}$ to define uniform priors for $\phi_j$ and $\phi_{0k}$, defined over 25\% of the image window \citep{banerjee2008gaussian}. Finally, we set $h_\alpha = 1$, which corresponds to an induced prior distribution for $\alpha_{jk}^2$ with a median of 0.45 and with 95\% of the central mass between 0.001 and 5.04, under the LGCP-N model. For the LGCP-BL model, however, the prior variance depends on both $\tau_{j k}^2$ and $h_{\alpha}$ and was adaptively regulated through the tuning parameter $\lambda^2$, which was assigned a hyperprior of $\textrm{Gamma}(0.7, 0.7)$.

In the analysis, we used $M$=$N$=$64$ (see Figure \ref{fig:images_steps}c) and $M_e$=$N_e$=$128$ for the extended grid in circulant embedding. For each dataset, we ran an MCMC chain for 500,000 iterations, discarding the first half as burn-in. Convergence of the Markov chains was assessed through visual inspection of trace plots and evaluation of joint posterior probability values over the course of the MCMC run, following the approach in \cite{diggle2013spatial}. Model selection and comparison were conducted by calculating the DIC described in Section \ref{sec:dic}.

\begin{figure}[ht]
\centering
\includegraphics[width=16cm]{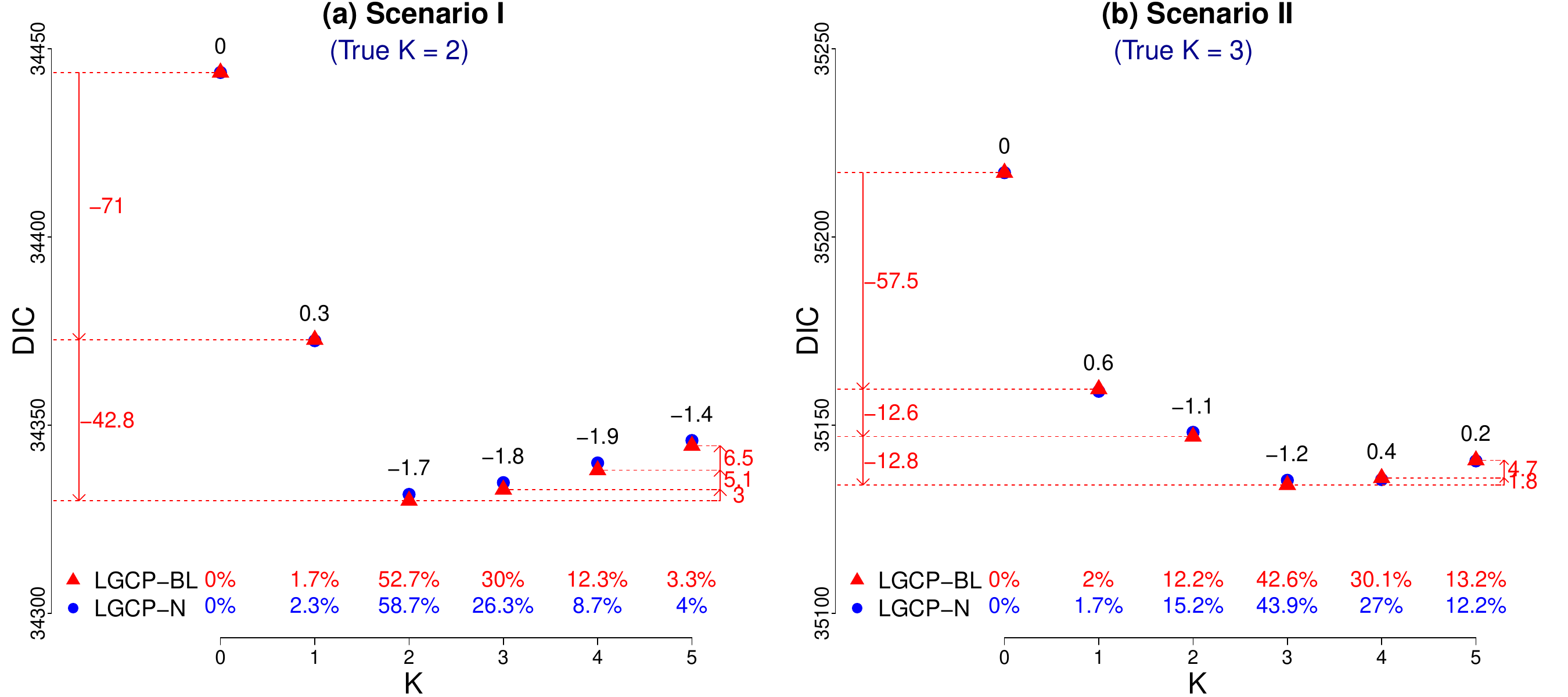}
\caption{Model evaluation based on the deviance information criterion (DIC) in simulation studies: The differences in mean DIC between LGCP-BL models with $K$ latent processes, DIC$_{BL}(K)$-DIC$_{BL}(K-1)$, are shown as numeric values in red. The mean differences in DIC between LGCP-BL and LGCP-N models for each $K$, DIC$_{N}(K)$-DIC$_{BL}(K)$, are provided as numeric values in black. The percentages of selected $K$ for each model across the 300 simulated datasets are presented at the bottom.} \label{fig:sim_dic}
\end{figure}

\subsection{Results of simulation studies}

For each simulated dataset, we focused on the model with $K$ selected by the DIC and summarized the estimated quantities across 300 datasets for each scenario. Since the results between the two proposed models were largely similar, our presentation primarily focuses on LGCP-BL unless stated otherwise.

In Scenario I, the DIC selected the LGCP-BL model with the true $K$ in 53\% of the simulations and models with $K$ differing by at most 1 from the true value ($K$=1, 2, 3) in 84\% of the simulations (Figure \ref{fig:sim_dic}). In comparison, under the same setting, the cross-validation approach used in the frequentist framework selected the true model in 41\% of the simulations and models with $K$ within $\pm 1$ of the true value in 67\% of the simulations (see Appendix of \cite{waagepetersen2016analysis}), highlighting the advantage of DIC-based model selection in the Bayesian framework. In Scenario II, the DIC selected the true model in 43\% of the simulations, and this percentage increased to 85\% when models with $K$ within $\pm 1$ of the true value were included. When the true models were not selected, the DIC criterion tended to favor models with higher values of $K$. For example, LGCP-BL models with $K$$<$ 2 were selected in only 1.7\% of simulations in Scenario I, and the models with $K$$<$ 3 were selected in 14.2\% of simulations in Scenario II.

In both scenarios, the DIC differences between models with $K$ and $K$$-$ 1 showed a large drop from $K$$=$ 0 to the true value of $K$, indicating substantial improvement in model fit (for LGCP-BL, $-114$ for Scenario I and $-83$ for Scenario II). For $K$ greater than the true value, the DIC showed a slight increase, suggesting potential overfitting with the inclusion of additional latent processes. While smaller for LGCP-BL in many cases, the differences in DIC between the two models were negligible ($<$2), based on the guideline outlined in Section \ref{sec:dic}.

The fitted models successfully capture the true trends of CCF and PV across distances (Figure \ref{fig:sim_ccf_pv_sel} and Supplementary Materials D). Note that because the results are based on models selected by DIC, they include models with varying numbers of shared latent processes (often more than necessary). The differences between LGCP-BL- and LGCP-N-based estimates are mostly negligible; however, LGCP-BL exhibits smaller variation in PV estimates for certain patterns (see Figure \ref{fig:sim_ccf_pv_sel}(n) and \ref{fig:sim_ccf_pv_sel}(o); Figure D.1(k) in Supplementary Materials).

\subsection{Comparison with a two-step frequentist approach}

We conducted an additional simulation study to compare the proposed Bayesian models with the two-step frequentist method of \citet{waagepetersen2016analysis}. The data were generated under the same setting as in their original study. Details of the simulation setup, implementation, and results are provided in Supplementary Materials E.

\section{Application to Microbiome Biofilm Image Data}

As outlined in Section \ref{sec:data}, we applied the proposed Bayesian framework to 100 image datasets to explore the multivariate dependence structure among the five taxa: \emph{Actinomyces}, \emph{Rothia}, \emph{Streptococcus mitis}, \emph{Streptococcus salivarius}, and \emph{Veillonella}.

\subsection{Specification of hyperparameters and analysis settings}

The black regions outside microbial consortia and the host epithelial cells, shown in white in Figure \ref{fig:images_steps}(a), were excluded from the analysis of each image. Microbial consortia on the tongue dorsum are densely organized, highly structured biofilms with well-defined perimeters and an epithelial core. The black regions lack microbial content or relevant host structures; including them would bias estimates of cross correlation among microbial taxa \citep{schillinger2012co}. Additionally, epithelial cells, displayed in white, serve as the structural core for microbial clustering and are not colonized internally by microbes; including them would therefore likewise cause an over-estimate of microbial cross correlation. Therefore, both regions were excluded from the spatial domain of our analyses using indicator variables on the computational grid, as represented by the black areas in Figure \ref{fig:images_steps}(c).

We considered two different analyses. In Analysis I, the hyperparameters for both models were set to the same values as described in Section \ref{sec:sim_hyperSetting}. In Analysis II, as part of a sensitivity analysis, the value of $h_\alpha$ was changed to 100 for both LGCP-BL and LGCP-N. All other analysis settings, including the MCMC run, grid specification, and convergence assessment, followed those outlined in Section \ref{sec:sim_hyperSetting}. The image-specific results were synthesized using the Bayesian hierarchical modeling approach described in Section \ref{sec:pooled}.

\begin{figure}[htp]
\centering
\includegraphics[width=16.5cm]{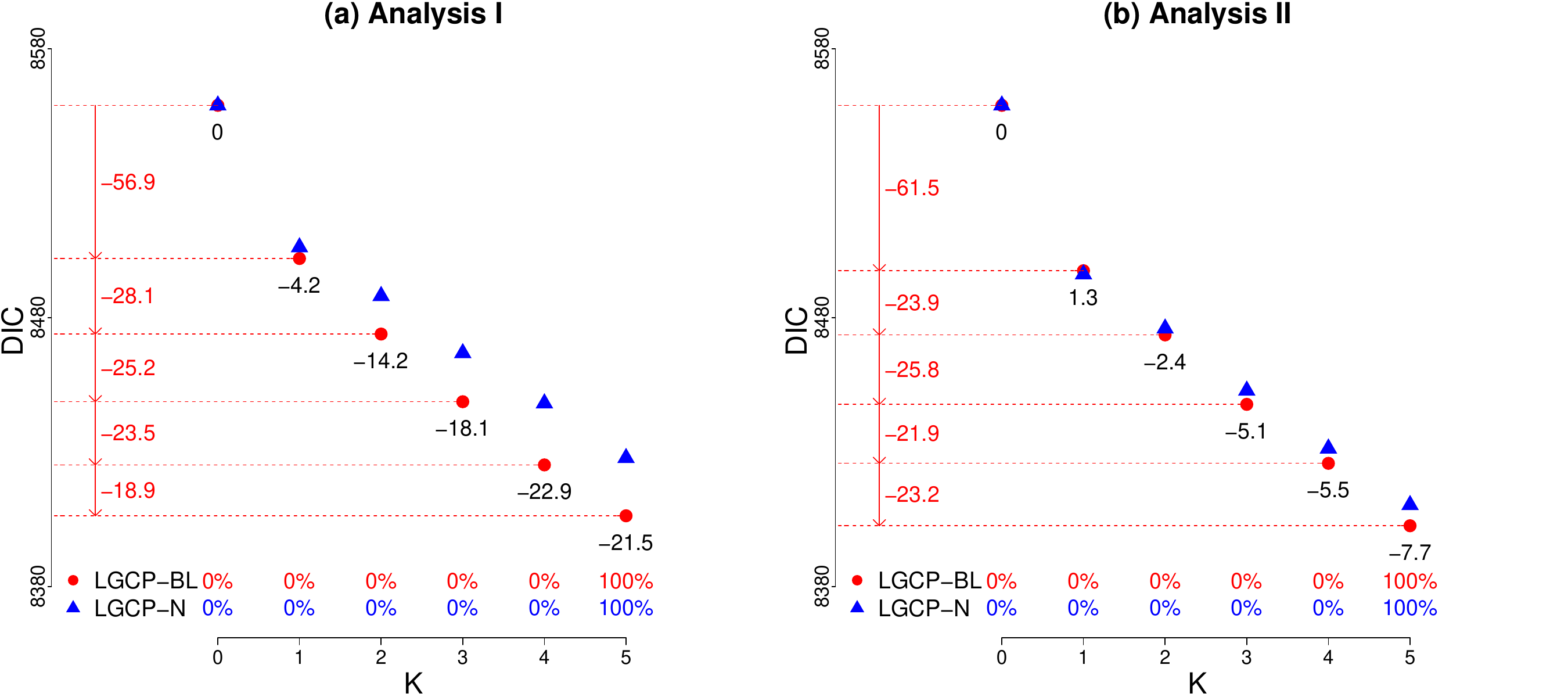}
\caption{Deviance information criterion (DIC)-based evaluation of models fitted to real biofilm image data: The differences in mean DIC between LGCP-BL models with $K$ latent processes, DIC$_{BL}(K)$-DIC$_{BL}(K-1)$, are shown as numeric values in red. The mean differences in DIC between LGCP-BL and LGCP-N models for each $K$, DIC$_{N}(K)$-DIC$_{BL}(K)$, are provided as numeric values in black. The percentages of selected $K$ for each model across the 100 images are presented at the bottom.} \label{fig:dataAll_dic}
\end{figure}

\begin{figure}[htp]
\centering
\includegraphics[width=15cm]{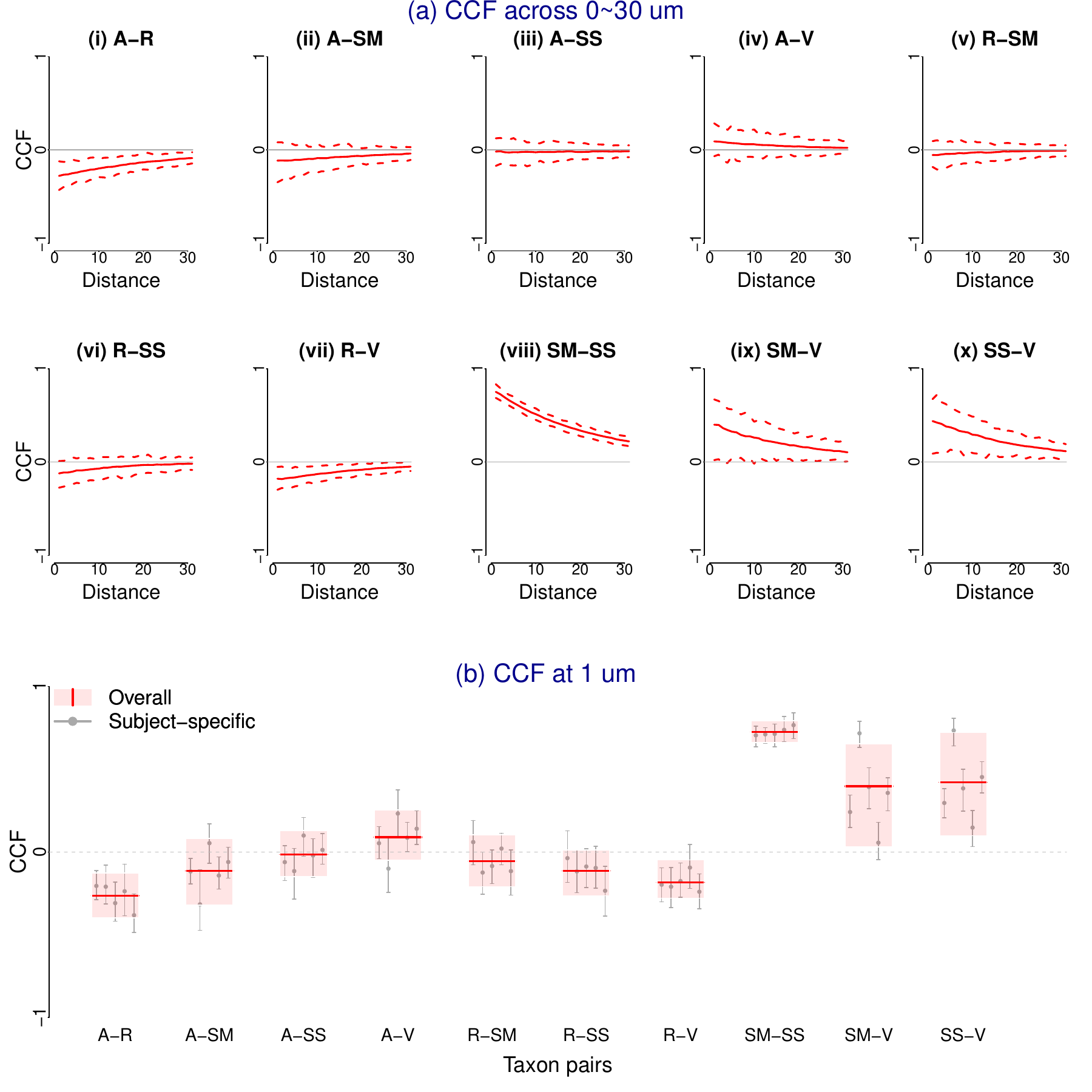}
\caption{Analysis I: Pooled analysis of cross correlation functions (CCFs) estimated using LGCP-BL for 100 human oral microbial biofilm image datasets from five participants: (a) The median (solid curves), 2.5\%, and 97.5\% (dashed curves) quantiles of the pointwise posterior medians for the global CCF across distances from 0 to 30 $\mu m$ for the 10 pairs of taxa. (b) The median, 2.5\%, and 97.5\% quantiles of the global (red line and shaded area, respectively) and donor-specific (grey point and line, respectively) CCF at 1 $\mu m$. A: \emph{Actinomyces}, R: \emph{Rothia}, SM: \emph{Streptococcus mitis}, SS: \emph{Streptococcus salivarius}, V: \emph{Veillonella}.} \label{fig:data1_ccf_combine_BL}
\end{figure}

\begin{figure}[htp]
\centering
\includegraphics[width=15cm]{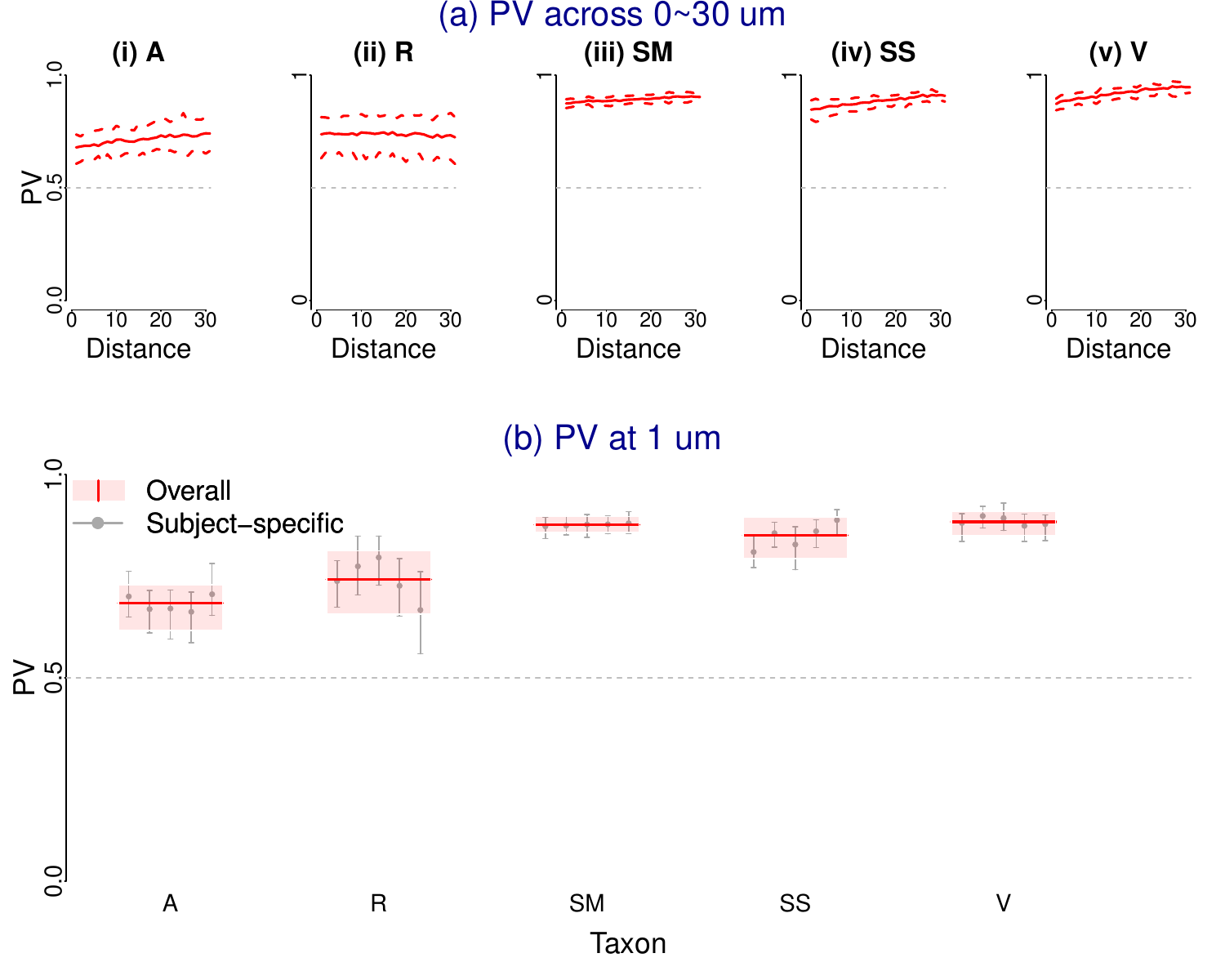}
\caption{Analysis I: Pooled analysis of proportion of variance functions (PVs) estimated using LGCP-BL for 100 human oral microbial biofilm image datasets: (a) The median (solid curves), 2.5\%, and 97.5\% (dashed curves) quantiles of the pointwise posterior medians for the global PV across distances from 0 to 30 $\mu m$ for five taxa. (b) The median, 2.5\%, and 97.5\% quantiles of the global (red line and shaded area, respectively) and donor-specific (grey point and line, respectively) CCF at 1 $\mu m$. A: \emph{Actinomyces}, R: \emph{Rothia}, SM: \emph{Streptococcus mitis}, SS: \emph{Streptococcus salivarius}, V: \emph{Veillonella}.} \label{fig:data1_pv_combine_BL}
\end{figure}

\subsection{Results of biofilm image data analysis} \label{sec:data_results}
In the main paper, we presented results corresponding to the LGCP-BL model under Analysis I. Additional results, including those from the LGCP-N model and from Analysis II, are provided in Supplementary Materials F. In Analysis I, the DIC favored models with $K$=$5$ across all 100 image datasets, providing strong support for the use of five latent processes to characterize the multivariate distribution of the five taxa in the biofilm data (Figure \ref{fig:dataAll_dic}). Notable DIC drops were observed as latent processes were added, from $K$=$0$ (null model) to $K$=$5$ (final selected model), with decreases ranging from $-56.9$ to $-18.9$. Adjusting $h_\alpha$ to 100 in Analysis II did not change this overall trend, as $K$=$5$ remained consistently selected for all datasets. In contrast to the simulation study results, the differences in DIC between LGCP-BL and LGCP-N were substantial. LGCP-BL consistently yielded much lower DIC values, ranging from $-22.9$ to $-2.4$, across all settings except at $K$=$1$ in Analysis II, where the difference was negligible \citep{spiegelhalter2002bayesian}. The five images, denoted as (Subject \#, image \#), with the largest drop in DIC between LGCP-BL and LGCP-N were (2, 8), (3, 4), (3, 19), and (5, 7), which contain relatively sparse taxa compared with other images (see Supplementary Materials A).

Three pairs of taxa, \emph{Streptococcus mitis}-\emph{Streptococcus salivarius}, \emph{Streptococcus mitis}-\emph{Veillonella}, and \emph{Streptococcus salivarius}-\emph{Veillonella}, exhibited clear positive spatial correlations. Two pairs of taxa, both involving \emph{Rothia}, showed slight negative correlations: \emph{Actinomyces}-\emph{Rothia} and \emph{Rothia}-\emph{Streptococcus salivarius}  (Figure \ref{fig:data1_ccf_combine_BL}). The large uncertainty observed in the subject-specific CCF estimates for pairs involving \emph{Actinomyces} in subject 2 might be attributed to the sparsity of this taxon in the images from that subject (Figure \ref{fig:data1_ccf_combine_BL}(b) and Supplementary Materials A). In contrast, the large uncertainty for pairs involving \emph{Rothia} in subject 5 were not due to taxon sparsity, as \emph{Rothia} is abundant in those images, and was primarily driven by the variability in the estimates.

Three taxa, \emph{Streptococcus mitis}, \emph{Streptococcus salivarius}, and \emph{Veillonella}, exhibited consistently high PV values ($>$0.8) across the range of inter-cellular distances, indicating that a large proportion of the variance in their spatial structure was attributed to shared latent processes (inter-taxon relationships)(Figure \ref{fig:data1_pv_combine_BL}). In contrast, \emph{Actinomyces} and \emph{Rothia} had estimated PV values between 0.6 and 0.8, demonstrating a smaller but still substantial contribution from inter-taxon relationships. In addition, there was greater variability in the subject-specific PV estimates at 1 $\mu m$ for \emph{Actinomyces}, \emph{Rothia}, and \emph{Streptococcus salivarius} compared with the other two taxa (Figure \ref{fig:data1_pv_combine_BL}(b)).

The magnitude of these estimated spatial inter-taxon correlations (Figure \ref{fig:data1_ccf_combine_BL}) aligned with that reported in previous studies \citep{wilbert2020spatial} and with established principles of microbial community interactions. \emph{Actinomyces} is frequently observed near the epithelial core, while \emph{Rothia} typically forms a cortical layer around the consortium. This pattern was supported by the negative CCF observed for the \emph{Actinomyces}-\emph{Rothia} pair. The positive spatial correlation estimated for the \emph{Streptococcus mitis}-\emph{Streptococcus salivarius} pair likely arises from shared environmental preferences, metabolic cooperation, and synergistic biofilm formation \citep{senthil2024oral}. Similarly, the positive spatial correlation between each of the two \emph{Streptococcus} species and \emph{Veillonella} could be attributed to cooperative metabolic interactions. \emph{Veillonella} is known to use lactate produced by \emph{Streptococcus}, which may facilitate the two taxa's coexistence and coaggregation within oral biofilms \citep{ramsey2011metabolite, mclean2012identifying, mashima2015interaction, zhou2021veillonellae}.

\section{Discussion}

In this paper, we proposed a novel Bayesian framework for multivariate spatial point pattern analysis of imaging data. This new approach addresses key limitations of methods that are widely used for estimating co-occurrence or spatial clustering of cell types, including the LDA method as implemented in \texttt{daime}, and frequentist LGCP models. Our approach allows for the simultaneous analysis of more than two object types (e.g. cell types, gene signals, or taxa), enabling comprehensive quantification of multivariate dependence structures such as the CCF and PV. It also produces normalized and interpretable spatial correlation estimates, overcoming the restriction to positive real values inherent in non-parametric methods. Although we are not the first to propose models that allow for such rescaling, to our knowledge we are the first to demonstrate how they can be applied to FISH images of cells. This critical contribution paves the way for standardized, model-based metrics of spatial clustering with a range of (-1,1). We anticipate that as use of this standardized approach expands, biologists can begin to develop intuition for what different values might mean within specific cellular contexts. The proposed framework also avoids underestimating covariance structure uncertainty, a common issue in non-likelihood-based frequentist approaches for LGCP models, and provides direct uncertainty quantification through posterior distributions.

Several features of the proposed framework bear emphasizing: the DIC-based model selection approach, priors that can be tuned up or down in terms of sparsity or flexibility, and a method for pooling data across images. To allow a principled approach for model selection, we designed a DIC specifically for the proposed multivariate LGCP models. The DIC demonstrated superior performance in selecting the model that best or nearly best characterizes the underlying multivariate dependence structure among multi-type point patterns. To afford the user the ability to tune the complexity (and flexibility) of the CCF, we considered two priors for factor loading parameters that control the degree of contribution of shared latent processes to the type-specific intensity processes. The model with lasso shrinkage priors exhibited marginal improvements in model fit in simulation studies and substantially better fit when applied to real microbiome biofilm image data. To effectively pool and synthesize model results from the set of image-specific analyses, we developed a Bayesian hierarchical modeling approach. This last feature offers one of the most important practical benefits of the proposed modeling framework, as it enables exploration of subject-specific variation in the CCF and PV, along with model-based estimates of their variability. 

One notable challenge of the proposed method is its computational speed for performing MCMC sampling in high-dimensional models. While fitting models to a set of images can be parallelized across images in a computing cluster, each MCMC run itself cannot be parallelized. Two possible approaches could greatly reduce computation time. One possibility is to use discrete uniform priors for spatial range parameters and precomputing key operations for high-dimensional spatial covariance matrices within the prior domain. Another possibility is to use integrated nested Laplace approximation or other approximate Bayesian methods \citep{teng2017bayesian, flagg2023integrated}.

The choice of computational grid resolution is not straightforward and requires careful consideration. For most environmental datasets, a grid naturally accompanies the data, as these are often products of satellites or forecast systems. However, this is not the case for count data, such as in our application. The resolution of the computational grid must balance estimation accuracy with computational complexity \citep{diggle2013spatial}. However, it is important to note that a finer grid does not always lead to better approximations of continuous processes. For instance, previous research has demonstrated that, under infill asymptotics, increasing the number of observations within a fixed domain can result in inconsistent estimates of range and variance parameters for Gaussian processes with Mat\'{e}rn covariance functions \citep{zhang2004inconsistent, zhang2005towards}. Moreover, challenges related to grid resolution arise in the context of spatial range parameters. In geostatistical applications with point-referenced data, variograms are often used as a tool to estimate spatial range parameters and inform grid size selection \citep{taylor2015bayesian}. However, this approach is not feasible for biofilm image data, or point process data in general, due to the inherent differences in data structure and scale. Consequently, there is limited literature addressing quantitative methods to optimize grid resolution, highlighting this as an important area for future research.

In numerical studies, use of a DIC metric for model selection was effective and outperformed cross-validation methods typically used in the conventional frequentist approaches. However, this process required fitting all candidate models to evaluate model fit, which can be computationally expensive. In contrast, regularized frequentist approaches offer automated selection of the number of latent processes \citep{choiruddin2020regularized}. This is mainly because Gaussian and Bayesian Lasso priors for $\bfalpha$ used in our framework provide continuous shrinkage toward zero but do not induce exact sparsity, and thus they are not directly applicable to threshold-based model selection. In the Bayesian paradigm, more efficient model selection metrics---for example, those building upon sparsity-inducing priors or post hoc thresholding rules and advanced computational schemes such as birth-death processes and Bayesian model averaging---may help streamline the selection process while maintaining the rigor of the Bayesian framework.

The proposed framework adopts exponential correlation functions for both shared and type-specific latent processes due to their simplicity, interpretability, and computational stability. While this is a common choice in multivariate spatial models, we acknowledge that alternative covariance functions, such as the Mat\'ern class with estimable smoothness parameters, may offer additional flexibility in capturing a wider range of spatial dependence structures. These alternatives can potentially be assessed and compared using model selection criteria like the DIC.

Regarding interpretability of the latent structure, while the latent Gaussian processes themselves are not directly visualized in our results, their influence is conveyed through estimated loadings ($\bfalpha$) and interpretable summaries such as the CCF and PV. These quantities reflect how taxa are interconnected via the shared latent structure. This approach aligns with previous multivariate LGCP frameworks (e.g., \cite{waagepetersen2016analysis}), where interpretability was provided through schematic model representations rather than by direct visualization of latent fields. Low-rank projections or spatial summaries of the latent surfaces may provide additional interpretability in applied contexts.

In the simulation results, the estimated PV curve for Type 2 exhibits a consistent upward bias. This behavior appears to stem from a combination of factors. First, the true PV for Type 2 decreases over distance, making it more difficult to estimate than the increasing trends observed in other types. Second, the PV depends solely on a shared latent process $U_{01}$ that also exclusively contributes to the PV for Type 1 (with $\alpha_{12} = \alpha_{22} = 0$), introducing potential cross-type borrowing. Third, the signal strength is relatively weak, with true PV values remaining close to 0.5 across distances, where shared and type-specific contributions are more balanced and less clearly separable. The estimation challenges in this setting are reflected in the PV curve's wider 95\% credible intervals for for Type 2. Nevertheless, these intervals consistently include the true PV values, indicating that the Bayesian models provide reliable uncertainty quantification even under structurally challenging conditions. Future work could explore model extensions that incorporate adaptive priors or structured regularization to more flexibly distinguish variance sources in types with intermediate-to-weak signal or ambiguous decomposition patterns.

The proposed framework provides a strong foundation that can be adopted and extended to address additional scientific questions and improve its applicability. We are working to extend multivariate point process models to incorporate subject- and image-specific ``non-spatial" covariates. Generalization of analysis results in this work is constrained by the inclusion of only five healthy subjects, though microbial communities on the tongue dorsum are likely influenced by individual factors such as diet, oral hygiene, and overall health. We observed substantial variation in subject-specific CCF estimates for certain taxon pairs and in PV estimates for some taxa. By applying the extended framework capable of incorporating multi-level non-spatial covariates in data from a well-designed and larger study, we could more effectively identify sources of variation and potential associations with biofilm spatial structure.

The methods can also be generalized at the image level. Recent advancements in multivariate cluster point process models have been applied to quantify multi-taxa arrangements in microbiome biofilm image data from dental plaque samples \citep{majumder2024multivariate}. However, these methods require the pre-specification of which candidate taxa form multi-layered clusters and the assumption that all other taxa follow independent homogeneous processes. Integrating the multivariate LGCP into cluster point process models will enhance their flexibility and relax these restrictive assumptions, enabling more robust analyses of spatial patterns in microbiome data.

In conclusion, the proposed Bayesian framework for multivariate point pattern analysis offers researchers a robust and flexible statistical tool to identify and quantify complex multivariate dependence structures observed in microbiome biofilm images. By addressing key limitations of existing methods, providing straightforward uncertainty quantification, and enabling simultaneous analysis of multiple taxa, this framework advances the study of spatial patterns in microbial communities or similar biomedical image data. The methods developed in this research, combined with publicly available software and insights from our numerical studies, have the potential to benefit a wide range of disciplines beyond microbial ecology, particularly in applications involving multi-type point patterns and spatial dependencies.

\section*{Funding}

This work was supported by the National Institute of Dental and Craniofacial Research (R21DE026872) and the National Institute of General Medical Sciences (R01GM126257).

\section*{Conflict of Interest}
None declared.

\bibliographystyle{apalike}
\bibliography{mlgcp}

\end{document}